\documentclass[10pt,notitlepage,aps,pra,showpacs,superscriptaddress,floatfix]{revtex4-1}

\usepackage[colorlinks=true,citecolor=blue,urlcolor=blue]{hyperref}
\usepackage[utf8]{inputenc}
\usepackage[T1]{fontenc}
\usepackage[sc,osf]{mathpazo}
\usepackage{amsmath, amsthm, amssymb,amsfonts}
\usepackage{graphicx}
\usepackage{dcolumn}
\usepackage{bm}
\usepackage{bbm}
\usepackage{mathtools}
\usepackage{comment}
\usepackage{color}
\usepackage{natbib}

\usepackage{graphicx}
\usepackage{amssymb}
\usepackage{amsmath}
\usepackage{amsthm}
\usepackage{color}
\usepackage{psfrag}
\usepackage{epsfig}
\usepackage{bbm}
\usepackage{bm}
\hypersetup{breaklinks}

\usepackage{soul}

\newcommand{\ket}[1]{| #1 \rangle}
\newcommand{\bra}[1]{\langle #1 |}

\newcommand{\beq}{\begin{eqnarray}}
\newcommand{\eeq}{\end{eqnarray}}

\newcommand{\ketbra}[1]{\ensuremath{| #1 \rangle \!\langle #1 |}}

\newcommand{\LL}{\mathcal{L}}
\newcommand{\WW}{\mathcal{W}}

\newcommand{\HH}{\mathcal{H}}

\newcommand{\Ain}{{\rm A_{\rm I}}}
\newcommand{\Aou}{{\rm A_{\rm O}}}
\newcommand{\Bin}{{\rm B_{\rm I}}}
\newcommand{\Bou}{{\rm B_{\rm O}}}

\newcommand{\M}{\mathcal M}

\newtheorem{theorem}{Theorem}
\newtheorem*{theorem*}{Theorem}
\newtheorem{lemma}{Lemma}

\DeclareMathOperator{\tr}{tr}

\begin{document}
\title{Composition rules for quantum processes: a no-go theorem}

\author{Philippe Allard Gu\'{e}rin}
\affiliation{Faculty of Physics, University of Vienna, Boltzmanngasse 5, 1090 Vienna, Austria}
\affiliation{Institute for Quantum Optics and Quantum Information (IQOQI), Austrian Academy of Sciences, Boltzmanngasse 3, 1090 Vienna, Austria}

\author{Marius Krumm}
\affiliation{Faculty of Physics, University of Vienna, Boltzmanngasse 5, 1090 Vienna, Austria}
\affiliation{Institute for Quantum Optics and Quantum Information (IQOQI), Austrian Academy of Sciences, Boltzmanngasse 3, 1090 Vienna, Austria}

\author{Costantino Budroni}
\affiliation{Institute for Quantum Optics and Quantum Information (IQOQI), Austrian Academy of Sciences, Boltzmanngasse 3, 1090 Vienna, Austria}

\author{\v{C}aslav Brukner}
\affiliation{Faculty of Physics, University of Vienna, Boltzmanngasse 5, 1090 Vienna, Austria}
\affiliation{Institute for Quantum Optics and Quantum Information (IQOQI), Austrian Academy of Sciences, Boltzmanngasse 3, 1090 Vienna, Austria}

\date{\today}

\begin{abstract}
A quantum process encodes the causal structure that relates quantum operations performed in local laboratories. The process matrix formalism includes as special cases quantum mechanics on a fixed background space-time, but also allows for more general causal structures. Motivated by the interpretation of processes as a resource for quantum information processing shared by two (or more) parties, with advantages recently demonstrated both for computation and communication tasks, we investigate the notion of composition of processes. We show that under very basic assumptions such a composition rule does not exist. While the availability of multiple independent copies of a resource, e.g. quantum states or channels, is the starting point for defining information-theoretic notions such as entropy (both in classical and quantum Shannon theory), our no-go result means that a Shannon theory of general quantum processes will not possess a natural rule for the composition of resources.


\end{abstract}
\maketitle

\section{Introduction}
Experimental tests with elementary quantum systems, most notably Bell tests, radically 
challenge the very notions of physical reality and cause-effect relations~\cite{Bell1964,FineTuningBellSpekkens}. Notwithstanding such fundamental 
novel effects, quantum mechanics still assumes a definite causal order of events. Namely, given two events, i.e. two operations performed 
locally in two quantum laboratories, say ${\rm A}$ and ${\rm B}$, we always assume that they are either time-like 
separated, hence, ${\rm A}$ cannot signal to ${\rm B}$ or vice versa, or they are space-like separated, hence, they cannot signal in either direction.

Motivated by the problem of quantum gravity, operational formalisms have been proposed for computing the joint probabilities for the outcome of local experiments, without the assumption of a fixed space-time background~\cite{Hardy2005, Hardy2007, Hardy2009, Oreshkov2012, Chiribella2013, Oeckl2016}. \textit{Process matrices}~\cite{Oreshkov2012} are introduced as the most general class of multilinear mappings of local quantum operations into probability distributions. The process matrix formalism provides a unified description of causally ordered quantum mechanics (quantum states and quantum channels), but also includes experimentally relevant non-causal processes such as the quantum switch~\cite{Chiribella2013, Araujo2015, OpticalQuantumSwitchProcopio, OpticalQuantumSwitchRubino, Goswami2018, Oreshkov2018, Oreshkov2016}. Furthermore, the formalism predicts novel and potentially observable phenomena, such as the violation of so-called {\it causal inequalities}~\cite{Oreshkov2012,Branciard:2016aa, Abbott:2016aa,Miklin_ineq:2017, Oreshkov2016}.

Moreover, it has been proven that such processes are able to provide advantages for quantum information processing tasks, both for computation and communication~\cite{Chiribella2013, AraujoPRL2014, Feix2015, AllardPRL2016, Baumeler2017, Araujo2017_CTC, Baumeler2018, Ebler2018}. One would, then, expect that a theory of information can be developed also for processes. Such a theory would deal with, e.g., rates of information compression and communication, i.e., a process-analog of the classical and quantum Shannon theory. 
A fundamental assumption in classical and quantum Shannon theory~\cite{Nielsen2002, Wilde2017} is the availability of multiple independent copies of a resource (for example a classical source of random variables, a quantum state, or a channel), which is at the basis of the definition of information-theoretic entropy, i.e., Shannon or von Neumann entropy. To be more concrete, in the example of Schumacher's compression~\cite{SchumacherCompression,Nielsen2002}, the optimal data compression of $n$ samples of an independent and identically distributed quantum source $\rho$ into $nS(\rho)+\delta$ qubits (with $\delta\rightarrow 0$ for $n\rightarrow \infty$), and the subsequent transmission, can be achieved only if the sender can act globally on multiple copies of the quantum state in which the information is encoded.

A natural question then arises, namely, whether a process matrix can be understood as a resource available in multiple (possibly identical) copies to experimenters, similarly to the example of Schumacher's compression above. Answering this question will provide us with deeper insight into the nature of process matrices. For instance, if we consider an experimental realization of a process, e.g., consisting of a sequence of optical elements as in photonic experiments~{\cite{OpticalQuantumSwitchProcopio,OpticalQuantumSwitchRubino}, one can easily imagine that it is possible to create two identical copies of the setup, and share them among the two parties. 
Alternatively, if one imagines that a process matrix does not only represent an experimental setup, but also the space-time structure~\cite{ZychThesis, CommonDirectCauseSuperposition,GravityQuantumSwitch}, then it is harder to imagine how two ``copies of spacetime'' may be shared between the two parties. More generally, such a composition rule should not be only about identical copies, but it should also allows us to combine different processes. 

It is important, at this point, to distinguish two different scenarios and their corresponding composition rules. On the one hand, one may simply ask what is the rule for composing different processes \textit{independently}, with the requirement that experimenters act locally on each copy of the process; this rule is given by the tensor product. On the other hand, going back to the example of Schumacher's compression protocol, one may require that a single experimenter (or many experimenters for multipartite systems) has access to multiple copies of a process, in order to perform a protocol that involves global operations. We will see that the latter notion is incompatible with the definition of a process.

For quantum states, quantum channels, or for any collection of processes with the same definite causal order~\cite{Chiribella2009, QuantumCausalModelsCosta}, the parallel composition can be described by the tensor product. However, it is known that a parallel composition of process matrices via the tensor product can fail~\cite{Jia2018}, as the resulting process matrix contains causal ``double-loops''~\cite{Oreshkov2012}, which give rise to the ``grandfather paradox'', or equivalently, to unormalised probabilites. 

In this work, we show that under weak assumptions (bilinearity, every output is a valid process matrix, reduction to the usual tensor product for definite causal structure) there exists no composition that allows the experimenters to have access to multiple shared processes. This result means that many information theoretic protocols relying on many copies of a resource have no straightforward generalization to process matrices.

\section{Preliminary notions}
The most general operation that can be performed on a quantum system is represented by a quantum instrument, namely, a collection $\{ \M_{a}\}_a$ of completely positive trace-nonincreasing maps
that sum up to a trace-preserving map $\M=\sum_a \M_a$. An operation represented by the instrument $\{ \M_{a}\}_a$ will give an output $a$ with probability $P(a)=\tr[\M_a(\rho)]$ and transformation of the state $\rho \mapsto \M_a(\rho)/P(a)$. We admit the possibility of an input $x$, and label the corresponding operations as $\{\M_{a|x}\}_{a,x}$. Such maps can be represented as matrices via the Choi-Jamio\l{}kowski isomorphism~\cite{Jamiolkovski,Choi}
\begin{equation}\label{eq:choim}
\M_{a|x} \mapsto M_{a|x} = \sum_{ij} |i \rangle \langle j|^{A_I} \otimes \M_{a|x}(|i \rangle \langle j|)^{A_O}.
\end{equation}
We will call $M_{a|x}$ the Choi matrix of $\M_{a|x}$ \footnote{Alternatively, one could define $\M_{a|x}$ with a global transposition $^t$, taken w.r.t. the $\{\ket{ij}\}_{ij}$ basis, as in Ref.~\cite{Araujo2015}. This allow one to write the process matrix associated to a quantum state $\rho$ shared between the parties simply as $\rho_I\otimes \openone_O$, instead of $\rho^t_I\otimes \openone_O$. We will not use this convention here.}.
\begin{figure}[t]
\includegraphics[width=5cm]{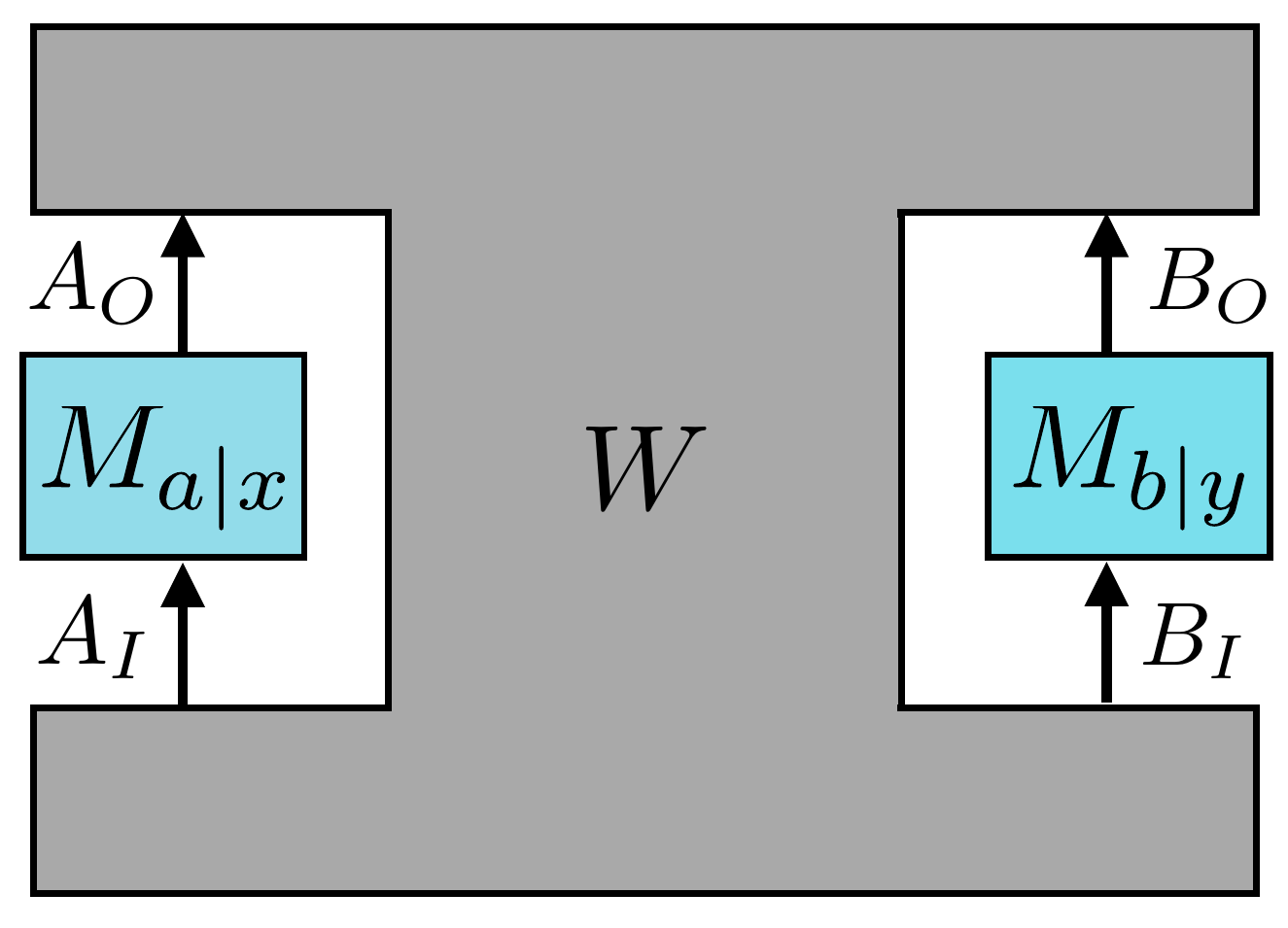}
\caption{Graphical representation of the probability rule Eq.~\eqref{eq:pabdef}.}
\centering
\label{fig:process}
\end{figure}
Consider a set of local operations, i.e., Choi matrices, $\{M_{a|x}^{\rm A}\}_{a,x}$ and  $\{ M_{b|y}^{\rm B}\}_{b,y}$, associated with Alice's and Bob's laboratories, where ${\rm A}$ denotes Alice's input-output space $\mathcal{H}_{\Ain}\otimes \mathcal{H}_{\Aou}$, and similarly for ${\rm B}$. A process $W$ is understood as the most general linear mapping of such operations into probabilities, which can be represented using the trace inner product as
\begin{equation}\label{eq:pabdef}
 p(ab|xy) =  \ \tr \left[ \Big(M^{\rm A}_{a|x}\otimes M^{\rm B}_{b|y} \Big) W^T \right],
\end{equation}
where $T$ denotes the transposition in the computational basis. A visual representation of this probability rule is given in Figure~\ref{fig:process}. In order to obtain valid probabilities, i.e., non-negative numbers summing up to one,  for arbitrary operations $\{M_{a|x}^{\rm A}\}_{a,x}$, $\{ M_{b|y}^{\rm B}\}_{b,y}$ (including operations that involve shared entangled ancillary systems), it can be proven~\cite{Araujo2015} that the following constraints must be satisfied
\begin{gather}\label{eq:bip1}
 W \ge 0 \, , \\
\label{eq:bip2} \tr W = d_O = d_{\Aou} d_{\Bou}, \\
\label{eq:bip3} {}_{\Bin \Bou}W = {}_{\Aou \Bin \Bou}W ,  \\
\label{eq:bip4} {}_{\Ain \Aou}W = {}_{\Ain \Aou \Bou}W  ,  \\
 \label{eq:bip5} W = {}_{\Bou}W + {}_{\Aou}W - {}_{\Aou \Bou}W , 
\end{gather}
where ${}_{X}W := \frac{\openone^{X}}{d_X} \otimes \tr_X W$.
The linear constraints in Eqs.~\eqref{eq:bip2}-\eqref{eq:bip5} can be written in a more compact form as
\begin{equation}
L_V(W)=W,
\label{eq:L_V}
\end{equation}
where $L_V$ is the projector onto the subspace of operators in $\LL(\HH_{\rm AB})$ that satisfy  Eqs.~\eqref{eq:bip3}-\eqref{eq:bip5}. We will denote such a linear subspace as $L_V(\LL(\HH_{\rm AB}))$. This projector enforces the normalisation of probabilites, and can be interpreted as preventing the paradoxes that would occur in processes with ``causal loops''~\cite{Oreshkov2012}. It is also convenient to define $\WW \subset \LL(\HH_{\rm AB})$ as the set of matrices that satisfy conditions in Eqs.~\eqref{eq:bip1}-\eqref{eq:bip5}, and similarly $\WW'$ for the spaces ${\rm A'B'}:=\HH_{\Ain '}\otimes \HH_{\Aou '}\otimes \HH_{\Bin '}\otimes \HH_{\Bou '}$.  If ${}_\Bou W = W$, one can show that Bob cannot signal to Alice, i.e., $p(a|x,y)=p(a|x,y')$ for all $a,x,y,y'$, we denote it as ${\rm A \preceq B}$ and we say that the process is causally ordered~\cite{Araujo2015}. Similarly, the case ${}_\Aou W = W$ correspond to the opposite causal order and it is denoted as ${\rm B \preceq A}$. If ${}_{\Aou \Bou} W = W$ we have at the same time ${\rm A \preceq B}$ and ${\rm B \preceq A}$, then $W$ represents a bipartite quantum state and we have no-signaling in both directions.

Similarly, in the case of $N$ parties ${\rm A^1,\ldots,A^N}$, linear constraints can be written in the compact form~\cite{Araujo2015}
\begin{equation}
L_{V_N}(W):={}_{\left[1- \prod_{i=1}^N (1-\Aou^i+\Ain^i\Aou^i) +\prod_{i=1}^N \Ain^i\Aou^i \right]}W=W,
\end{equation}
where the index $i$ runs through the different parties. Notice that if $W=W^1\otimes W^2$, then the set $\{1,\ldots,N\}$ can be split as $\chi_1\cup \chi_2$, with $\chi_1\cap \chi_2=\emptyset$, where $\chi_k$ indexes the parties appearing in $W^k$. Then 
\begin{equation}\label{eq:constr_tens}
\begin{split}
L_{V_N}(W_1\otimes W_2)= &W_1\otimes W_2 \Leftrightarrow\\
{}_{\left[1- \prod_{i \in \chi_1} (1-\Aou^i+\Ain^i\Aou^i) +\prod_{i \in \chi_1} \Ain^i\Aou^i \right]} W_1 =W_1 
\text{ and }& {}_{\left[ 1- \prod_{i\in \chi_2} (1-\Aou^i+\Ain^i\Aou^i) + \prod_{i\in \chi_2} \Ain^i\Aou^i \right]} W_2=W_2.
\end{split}
\end{equation}

\subsection{Examples} The process matrix formalism allows one to treat quantum states, quantum channels, and even situations where the causal order is indefinite, in a unified way. For example, the process matrix associated to a quantum state $\rho$ can be described as a single party process matrix, as $W = \rho^{A_I} \otimes \openone^{A_O}$. The process matrix associated to $N$ spatially separated copies of the state is a $N$-partite process $W = \prod_{i=1}^N \rho^{A_I^i} \otimes \openone^{A_O^i}$, where each of the $A_I^i$ and $A_O^i$ are isomorphic. However, one could also consider the same $W$ as a global single party process,  with input Hilbert space $A_I = \prod_i A_I^i$, and output Hilbert space $A_O = \prod_i A_O^i$.

A quantum channel $\mathcal{C}: \LL( \HH_{A_O}) \to \LL( \HH_{B_I})$, connecting the output Hilbert space of Alice to Bob's input Hilbert space, can be described in process matrix language as $W = C^{A_O B_I }$, where $C$ is the Choi matrix of the channel $\mathcal{C}$, as defined by Eq.~\eqref{eq:choim}. The process matrix describing $N$ parallel uses of the channel $\mathcal{C}$ is simply $W = \prod_{i=1}^N C^{A_O^i B_I^i}$. Again, this process can be considered as a $2N$-partite process, or as a bipartite process with $A_O := \prod_i A_O^i$ and $B_I = \prod_i B_I^i$.

\section{Composition rules} From the above considerations, it seems that one could simply take the tensor product as a composition rule to obtain multipartite processes representing multiple independent copies of a resource. In fact, Eq.~\eqref{eq:constr_tens} implies that whenever the linear constraints are satisfied for both $W_1$ and $W_2$, then the corresponding multipartite constraints will be satisfied for $W_1\otimes W_2$. 

\begin{figure}[t]
\includegraphics[width=5cm]{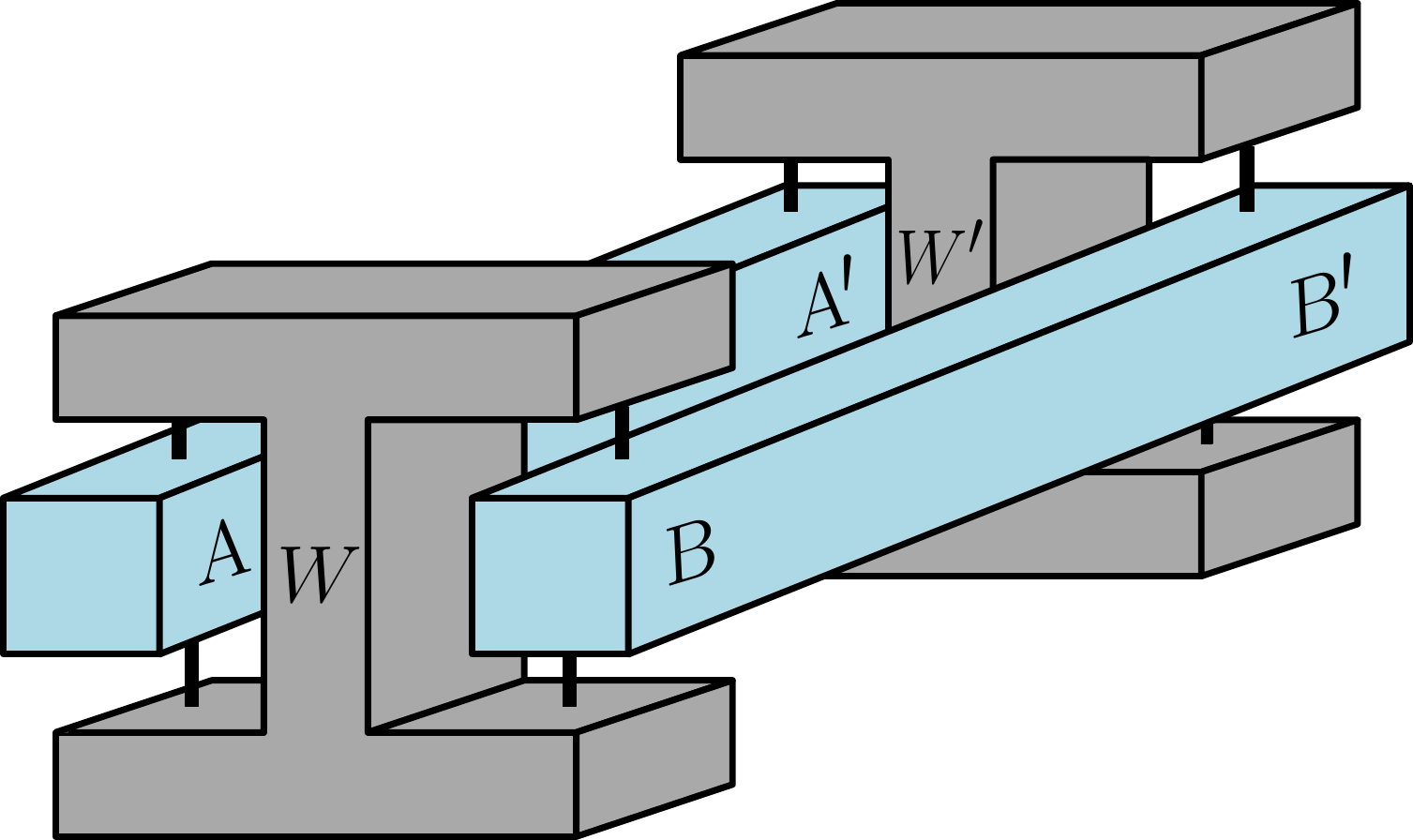}
\caption{The tensor product compostion rule $\mu(W, W') = W \otimes W'$. Here $A A'$ is a composite party that can perform general quantum operations $\mathcal{L}(\mathcal{H}_{A_I A'_I}) \to \mathcal{L}(\mathcal{H}_{A_O A'_O})$, and similarly for $B B'$; the corresponding probabilities are given by Eq.~\eqref{eq:pabdef}. We shall show that this composition rule does not satisfy all requirements that we demand on such a rule.}
\centering
\label{fig:tensor}
\end{figure}
The situation is different, however, if we require $W_1$ and $W_2$ to be shared by {\it the same} parties. To keep the discussion simple, consider only two parties, Alice and Bob, who share two possible processes, $W_1\in \WW$ and $W_2\in \WW$'. We want now to create the composite process $\mu(W^1,W^2)$ such that it is still a bipartite one, i.e., Alice can access both the systems $\Ain\Aou$ and $\Ain'\Aou'$, and Bob both $\Bin \Bou$ and $\Bin' \Bou'$. If both processes have the same definite order, i.e., ${}_\Aou W_1 = W_1$ and ${}_{\Aou'} W_2 = W_2$, or the analogous condition with $\Bou,\Bou'$, then, we know from standard quantum theory that the right operation for composing such processes is $W_1\otimes W_2$. This composition rule is represented in Fig.~\ref{fig:tensor}. One can easily prove that whenever the two processes do not have the same definite causal order, then $L_V(W^1\otimes W^2) \neq W^1\otimes W^2$, where $L_V$ is taken with respect to the bipartition $({\rm AA'},{\rm BB'})$~\cite{Jia2018}. For instance, consider the process
\begin{equation}\label{eq:pr_w}
W\otimes W, \text{ with } W= \frac{1}{2}( W^{\rm A \preceq B} +  W^{\rm B \preceq A})
\end{equation}
then, it is sufficient to check directly the violation of Eq.~\eqref{eq:bip5} with respect to the bipartition $({\rm AA'},{\rm BB'})$, namely, 
$W\otimes W \neq { {}_{\Bou\Bou'}(W\otimes W) } + { {}_{\Aou\Aou'}(W\otimes W) } - { {}_{\Aou \Aou' \Bou \Bou'}(W\otimes W) }$. This problem is illustrated in Figure~\ref{fig:process_forbidden}, where two processes $W, W'$ corresponding to channels in different directions can be seen to lead to a ``loop'', and to unnormalised probabilites. It is then natural to ask whether the tensor product can be replaced with another composition rule. 

\begin{figure}[t]
\includegraphics[width=7cm]{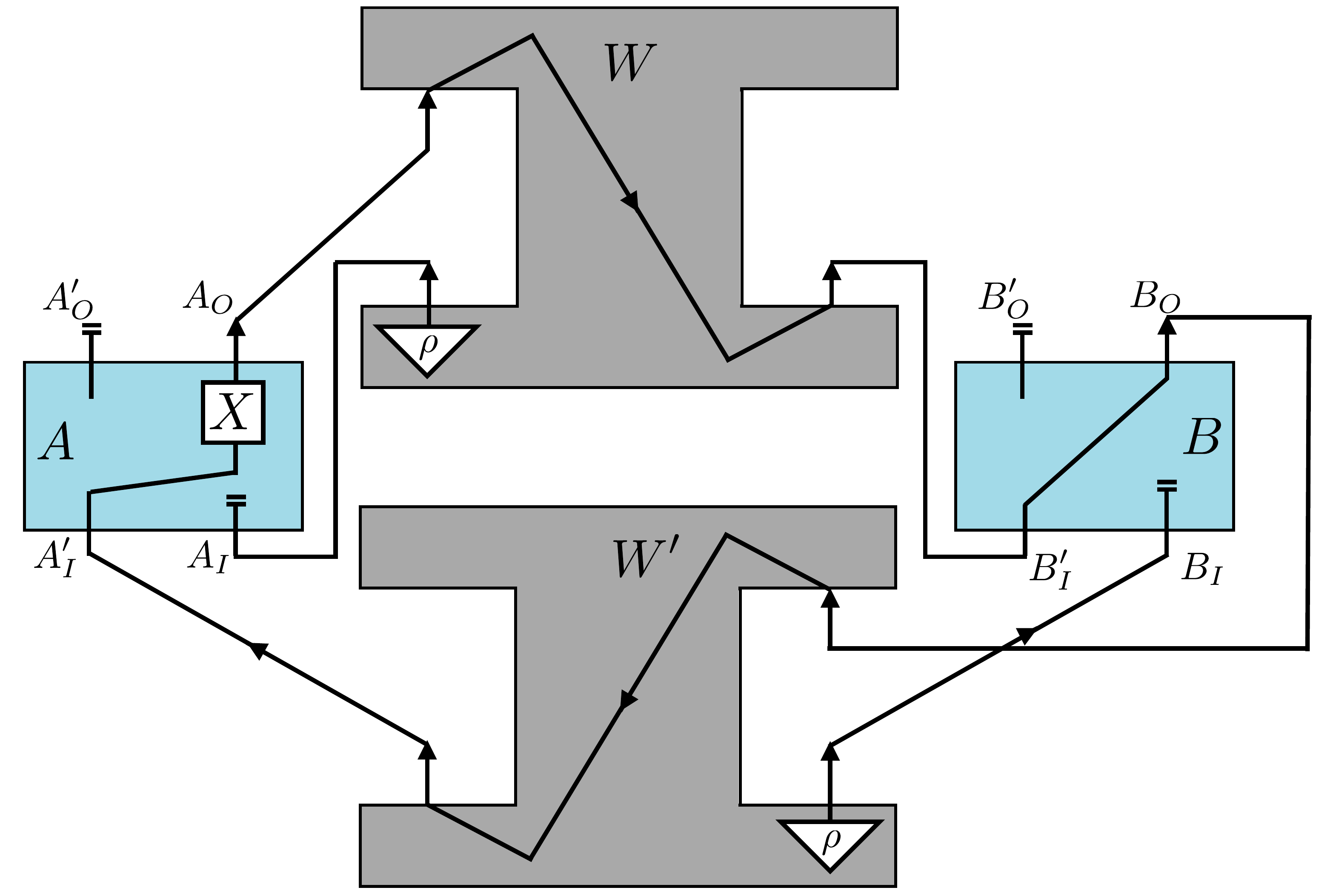}
\caption{The tensor product compostion rule $\mu(W, W') = W \otimes W'$ does not produce valid processes for all choices of $W$ and $W'$. Here the process $W$ corresponds to Alice recieving a state $\rho$, with an identity channel connecting her output system to Bob's input; $W'$ is the same thing with the order of the parties reversed. The specific choice of local maps ($X$ being the Pauli-$X$ unitary gate) have zero probability under the ``generalised Born rule'' Eq.~\eqref{eq:pabdef}, instead of one, as it should be for deterministic operations.}
\centering
\label{fig:process_forbidden}
\end{figure}

One may, however, argue that it is in principle possible to define more general composition rules that take this problem into account. For instance, one could take the tensor product and then ``project'' back the corresponding operator onto the space of valid process, or one could first decompose the process into a linear combination of processes in a definite order, then take the tensor product of each term and then recombine them. There are infinitely many possible recipes to define a composition rule; an abstract prescription for general composition rules is provided in Refs.~\cite{Perinotti2016, Bisio2018}. In the following, we will ask three reasonable and physically motivated requirements and show that there is no way of satisfying all three.

To define  our composition rule $\mu$, we may ask the following minimal requirements:
\begin{itemize}
\item[R.1] $\mu(W_1,W_2)$ is a valid process w.r.t. the bipartition $({\rm AA'},{\rm BB'})$, for  $W_1\in \WW,W_2 \in \WW'$ (Validity).
\item[R.2] $\mu(W_1,W_2)= W_1\otimes W_2$ if  $W_1\in \WW,W_2 \in \WW'$, and $W_1,W_2$ are in the same order, i.e.,  $({}_\Aou W_1 = W_1$ and ${}_{\Aou'} W_2 = W_2)$, or  $({}_\Bou W_1 = W_1$ and ${}_{\Bou'} W_2 = W_2)$ (Consistency).
\item[R.3]  $\mu(W_1,W_2)$ is convex linear in both arguments (Convex Linearity);
\end{itemize} 
Requirement R.1 is needed for the composition of two processes to still belong to a bipartite scenario, i.e., where Alice has access to both systems ${\rm AA'}$, and Bob to ${\rm BB'}$. R.2 is a consistency condition, i.e., the case of definite order should coincide with standard quantum theory. R.3 can be derived by requiring that our composition is well-behaved with respect to statistical mixtures, i.e., classical randomness, as explained in Appendix~\ref{app:A}.

It will be interesting to first consider a weaker assumption than R.1, because it will help us to single out the usual mathematical tensor product as a composition rule: 
\begin{itemize}
	\item[R'.1] $\mu(W_1,W_2)\geq 0$  for  $W_1\in \WW,W_2 \in \WW'$ (Positivity);
\end{itemize}

Assume that $\mu$ is a composition rule satisfying R'.1 (or R.1), R.2, R.3. Then there is a \textit{unique} real-linear extension $\mu^L$ that satisfies $\mu^L(W_1, W_2) = \mu (W_1, W_2)$, for all $W_1 \in \mathcal{W}, W_2 \in \mathcal{W}'$. 
By construction this extension satisfies:
\begin{itemize}
	\item[R'.3] $\mu(W_1,W_2)$ is real linear in both arguments (Linearity);
\end{itemize}	
For the linear extension, we only demand R.1' (or R.1) for process matrices as inputs, so it will trivially continue to be satisfied. As R.2 itself is a (bi)linear condition, the linear extension will satisfy it even when it is extended to the linear span of process matrices:
\begin{itemize}
	\item[R'.2] $\mu(W_1,W_2)= W_1\otimes W_2$ if  $W_1\in L_V(\mathcal L(\mathcal H_{AB})),W_2 \in L_V(\mathcal L(\mathcal H_{A'B'}))$, and $W_1,W_2$ satisfy $({}_\Aou W_1 = W_1$ and ${}_{\Aou'} W_2 = W_2)$, or  $({}_\Bou W_1 = W_1$ and ${}_{\Bou'} W_2 = W_2)$ (Consistency)
\end{itemize}
Details can be found in Appendix~\ref{app:A}. 

With our axioms, we will be able to prove
\begin{theorem}\label{th1}
The only function satisfying R'.1, R'.2, R'.3 is $\mu(W_1,W_2):= W_1\otimes W_2$.
\end{theorem}
\noindent{}
Theorem~\ref{th1} can be applied to the linear extension $\mu^L$, implying that $\mu(W_1 , W_2) = W_1 \otimes W_2$, and from that it will follow
\begin{theorem}\label{th2}
There exists no function satisfying R.1-R.3.
\end{theorem}
In particular, Th.~\ref{th1} will imply that for the multipartite case the choice of the composition rule is unique.
We will prove Th.~\ref{th1} for the simple case of local systems consisting of $n$-qubits, i.e., with local dimension $2^n$ for each one of $\Ain,\Ain',\ldots,\Bou,\Bou'$, the general proof can be found in Appendix~\ref{app:B}. Given Th.~\ref{th1}, for the proof of Th.~\ref{th2} it is sufficient to use the result of Ref.~\cite{Jia2018}, or the example in Eq.~\eqref{eq:pr_w}.

First, we need the following
\begin{lemma}\label{lemma:norm}
Given $A_1,A_2$ Hermitian operators such that $A_1\in L_V(\LL(\HH_{\rm AB}))$ and $A_2\in  L_V(\LL(\HH_{\rm A'B'}))$, and let $\mu$ be a composition rule satisfying R'.1-3. Then $\mu(A_1,A_2)=\mu(A_1,A_2)^\dagger$ and $\|\mu(A_1,A_2)\|\leq\| A_1\otimes A_2 \|$.
\end{lemma}

{\it Proof.--}For $A$ Hermitian, its norm can be written as: $\|A\| = \min \{ \lambda\ | -\lambda \openone \leq A \leq \lambda \openone\}$.
Consider $A_1\in L_V(\LL(\HH_{\rm AB}))$ and $A_2\in L_V(\LL(\HH_{\rm A'B'}))$ Hermitian and with  $\lambda_i= \|A_i\|$ for $i=1,2$. We define
\begin{equation}
\begin{split}
W_1^{\pm}=\lambda_1 \openone \pm A_1,\quad W_2^{\pm}=\lambda_2 \openone \pm A_2,
\end{split}
\end{equation}
which are valid processes, up to a normalization factor, on the spaces ${\rm AB}$ and ${\rm A'B'}$. We then have, 
\begin{equation}
\begin{split}
0 \leq \frac{\mu(W_1^+,W_2^+)+ \mu(W_1^-,W_2^-)}{2}
= \lambda_1 \lambda_2 \openone + \mu(A_1,A_2), \\
0 \leq \frac{\mu(W_1^+,W_2^-)+ \mu(W_1^-,W_2^+)}{2} 
= \lambda_1 \lambda_2 \openone - \mu(A_1,A_2),
\end{split}
\end{equation}
which implies $\mu (A_1,A_2) = \mu (A_1,A_2)^\dagger$ and $\|\mu(A_1,A_2)\|\leq \lambda_1\lambda_2$. In the above, we used R'.1 for  positivity, then R'.3 to split the different terms, and finally, R.2' to take the identity out of $\mu$. \qed

For the following, we need to specialize the form of the operator $A_1$ and $A_2$. We define the set of tensor products of either traceless operators or the identity as  
\begin{equation}
{\rm PTI}_{\rm AB} := \{ M=X^1_\Ain\otimes X^2_\Aou \otimes X^3_\Bin \otimes X^4_\Bou \ | M \in L_V(\LL(\HH_{\rm AB})) 
 ,\  X^i   \text{ identity or traceless }  \ \},
\end{equation}
and analogously for ${\rm A'B'}$. For $M\in {\rm PTI}_{\rm AB}$, an operator of the form $\openone + M$ is, up to normalization, a causally ordered process. With the above definition, we prove the following 
\begin{lemma}\label{lemma:pm1}
Let $\mu$ be a composition rule satisfying R'.1-3, and let $M\in {\rm PTI}_{\rm AB}$ and $N \in {\rm PTI}_{\rm A'B'}$ be Hermitian operators with eigenvalues in the interval $[-1,1]$. Given an eigenvector $\ket{k}$ of $M$ with eigenvalue $(-1)^k$ and an eigenvector $\ket{j}$ of $N$ with eigenvalue $(-1)^j$, we have
\begin{equation}
\mu(M,N)\ket{k,j}=(-1)^{k+j} \ket{k,j}
\end{equation}
\end{lemma}
{\it Proof.---}To prove the lemma, it is sufficient to consider the (unnormalized) processes $W_1^{k} := \openone + (-1)^{k+1} M$ and $W_2^{j} :=\openone + (-1)^{j+1} N$. By R'.2, $\mu(\openone,\openone)=\openone\otimes \openone$  and $\mu(M,\openone)=M\otimes \openone$, since for $M\in  {\rm PTI}_{\rm AB}$, either ${}_{\Aou}M=M$ or ${}_{\Bou}M=M$. Then,
\begin{equation}
\mu(W_1^k,W_2^j)=  \openone + (-1)^{k+1} M\otimes \openone + (-1)^{j+1} \openone \otimes N  + (-1)^{k+j} \mu(M,N).
\end{equation}
by R'.2 and R.3', and finally, by R'.1,
\begin{equation}
0 \leq \bra{k,j} \mu(W_1^{k},W_2^{j})\ket{k,j}
= 1-1-1+(-1)^{j+k}\bra{k,j} \mu(M,N)\ket{k,j},
\end{equation}
which implies
$\mu(M,N)\ket{k,j} = (-1)^{j+k}\ket{k,j}$,
since $\| \mu(M,N)\|\leq 1$, by Lemma~\ref{lemma:norm}. \qed

A straightforward corollary of Lemma~\ref{lemma:pm1} is that  $\mu(M,N)=M\otimes N$ whenever $M,N$ have eigenvalues only in $\{-1,1\}$. By linearity, this is enough to prove Th.~\ref{th1} for all processes defined on $n$-qubit systems (i.e., local dimension $2^n$) since we have a basis of operators, given by tensor products of Pauli matrices and the identity, that satisfy the assumptions. The same reasoning can be extended to arbitrary dimensions, see the details in Appendix~\ref{app:B}.

\section{ Discussion and conclusions}
In this letter, we considered the parallel composition of process matrices. As the tensor product is known to lead to invalid process matrices, we investigated whether there is another map that can describe this parallel composition. We only asked for three weak desiderata: First of all, in contrast to the usual tensor product, it should always result in a valid process matrix. Furthermore, it should reduce to the familiar tensor product in the case of definite causal order. At last, we demanded bilinearity for compatibility with the interpretation of convex mixtures as statistical mixtures. However we have seen that even those reasonable desiderata are incompatible with each other. 

Our results imply that an information theory of general quantum processes cannot rely on the assumption that multiple independent processes can be shared between two (or more) parties. In information theory, it is typical to assume that many independent samples of a random source, many independent uses of a channel, etc. are available, and that agents can perform global operations on many independent copies of the resource; this will not be possible in an information theory of general quantum processes. Rather, these results suggest that the proper setting for defining information-theoretic quantities such as entropies, capacities, etc., for process matrices is \textit{single-shot} information theory~\cite{Renner2006,Konig2009,Chiribella2016}.

One can infer from the main proof that even the case of two channels with opposing signalling direction will lead to a contradiction, which is perhaps unsurprising in the usual case of quantum mechanics on a fixed background spacetime. Indeed, suppose that an event $A$ is in the causal past of an event $B$, and that $A'$ is in the causal future of $B'$. Our desiderata that $A$ and $A'$ correspond to the same party can be interpreted as requiring that the events $A, A'$ occur at the same space-time point $p$. This could be the case, but then $B$ must be in the future light-cone of $p$, while $B'$ must be in it's past light-cone. It is thus impossible to satisfy the requirement that $B$ and $B'$ also occur at the same spacetime point.

Therefore any composition rule for process matrices must take care of removing the two-way signalling terms, whose impossibility has a clear interpretation as discussed above. We have shown that there is no linear way of doing so, if we ask for that our composition rule reduces to the usual tensor product in the case of two processes with the same definite causal order. 

However, there might exist reasonable non-linear composition rules, in the cases where processes have a concrete physical interpretation. A meaningful way to define an event for the composite party $A A'$ is by the ``simultaneous'' entering of both systems $\mathcal{H}_{A}$ and $\mathcal{H}_{A'}$ in a localised laboratory, and similarly for $B B'$. There can be a probability that the systems do not enter the laboratories simultaneously, in which case it is necessary to post-select on the runs of the experiment where this was indeed the case. Since the post-selection probability depends on the two processes that we wish to compose, the map will be non-linear. An important issue with such a post-selected composition map for information-theoretic applications is that the parallel composition of resources is usually a ``free operation'', while in the post-selected case it would have a probability of failure. 

\section{Acknowledgements}
We thank Ilya Kull, Ding Jia, Fabio Costa, Paolo Perinotti, and Lucien Hardy for useful discussions and comments. We acknowledge the support of the Austrian Science Fund (FWF) through the Doctoral Programme CoQuS, the projects I-2562-N27 and I-2906, the project M 2107 (Meitner-Programm) and the research platform TURIS, as well as support from the European Commission via Testing the Large-Scale Limit of Quantum Mechanics (TEQ) (No. 766900) project. P.A.G. acknowledges support from the Fonds de Recherche du Qu\'{e}bec -- Nature et Technologies (FRQNT). This publication was made possible through the support of a grant from the John Templeton Foundation. The opinions expressed in this publication are those of the authors and do not necessarily reflect the views of the John Templeton Foundation.



\providecommand{\href}[2]{#2}\begingroup\raggedright
\appendix

\begin{widetext}

\section{Linearity and convex linearity}\label{app:A}

In this appendix, we discuss convex-linearity and the linear extension of convex maps. First, let us argue why convex-linearity is a reasonable physical assumption. In operational approaches to physical theories~\cite{Hardy2001,Barrett2007}, one studies the probabilities that can be obtained from an abstract set of preparations and measurements. Given two preparations $\alpha$, and $\beta$, there exists an another preparation $\gamma$ that consists of preparing $\alpha$ with classical probability $p$, and preparing $\beta$ with probability $(1-p)$. The probability for any measurement on $\gamma$ is the weighted sum of the probabilities associated with preparations $\alpha$ and $\beta$. If we associate ``states'' with preparations, this means that the state space is convex linear. For example, the density matrix formalism can be seen to arise by adding classical uncertainty to the pure state formalism (i.e. kets in a Hilbert space). If one knows that with probability $p_j$, one prepares $\ket{j}$, then the density matrix is given by $\rho = \sum_j p_j \ket{j}\bra{j}$. Another motivation for allowing arbitrary probabilistic mixtures appears in Refs.~\cite{Hardy2009_foliable, Hardy2011}, where it is shown that it implies that optimal compression is equivalent to linear compression.

The same interpretation can be used for process matrices: if the process matrices $W_j$ are prepared with probabilities $p_j$, then all expectations values (and by that all statistics) can be calculated with the effective process matrix $W = \sum_j p_j W_j$. This can be seen by noting that $p(a,b) = \mathrm{Tr}[W M_a^{(A)} \otimes M_b^{(B)}]$ is a linear function in $W$ and applying the law of total probability.

Consistency demands that the composition rule $\mu$ remains compatible with this interpretation of convex mixtures: If the first process is $W_j$ with probability $p_j$ and the second process is $W_k'$ with probability $p_k'$, then the effective process matrices determining the statistics are $W = \sum_j p_j W_j$ and $W' = \sum_k p_k' W_k'$. The resulting combined process would be $\mu\Big( \sum_j p_j W_j, \sum_k p_k' W_k' \Big)$. However, a different point of view would be to say: With probabilities $p_j$ and $p'_k$ we combined the processes $W_j$ and $W'_k$ to $\mu(W_j, W'_k)$. So we prepared $\mu(W_j, W'_k)$ with probability $p_j p'_k$. Now, the effective process matrix is described by $\sum_{jk} p_j p'_k \mu(W_j, W'_k)$. As both points of view describe the same operational scenario, they have to be consistent:
\begin{align}
\mu\Big(\sum_j p_j W_j, \sum_k p_k' W_k'\Big) = \sum_{jk} p_j p'_k \mu(W_j, W'_k).
\end{align}

Next, we explain in further detail how to extend a function satisfying R.1 (or R'.1), R.2 and R.3 to a function satisfying R.1 (or R'.1), R'.2 and R'.3 on the linear span of all the process matrices.

Constructing the (bi)linear extension itself is a standard procedure in quantum information theory and is explained e.g. in Refs.~\cite{Hardy2001,Barrett2007} for general abstract state spaces. Let $S_1, S_2$ be two convex sets, and let $f: S_1 \to S_2$ be a convex linear map. Let $V_1, V_2$ be the real vector spaces obtained respectively from $S_1, S_2$ by taking their linear span. Then $f$ can be extended in the obvious way to a linear function $f^L: V_1 \to V_2$, defined by $f^L(\lambda a + b) = \lambda  f(a) +  f(b)$, for all $a,b \in S_1$, $\lambda \in \mathbb{R}$.

However, we still need to check that the bilinear extension still satisfies our postulates: We do not change R.1 (or R'.1), i.e. we only demand the output to be a process matrix (or positive) if the inputs are process matrices. Therefore R.1 (or R'.1) trivially continues to hold as the extension does not change the function on inputs that are process matrices.

Less trivial is how to generalize R.2. We will explicitly show that it still holds for the cases we need. Let us assume we have operators $M_1 \in L_V(\mathcal L (\mathcal H_{AB}))$ and $M_2 \in L_V(\mathcal L (\mathcal H_{A'B'}))$ with ${}_{A_O} M_1 = M_1$ and ${}_{A'_O} M_2 = M_2$ (or alternatively ${}_{B_O} M_1 = M_1$ and ${}_{B'_O} M_2 = M_2$). We now show that 
\begin{align}
\mu^L(M_1,M_2) = M_1 \otimes M_2.
\end{align}
By definition, $M_1$ and $M_2$ are allowed terms satisfying the projective condition~\eqref{eq:L_V}. Therefore there exist $\lambda_1,\lambda_2$ such that $\frac{\mathbb 1}{d_I}  + \lambda_1 M_1$ and $\frac{\mathbb 1}{d_{I'}} + \lambda_2 M_2$ are valid process matrices. Similarly $\frac{\mathbb 1}{d_I d_{I'}}$ itself is a valid process matrix, with no signaling at all. Using R.2 for the original $\mu$ on valid process matrices, we find for the linear extension:
\begin{align}
\mu^L\left(\frac{\mathbb 1}{d_I} + \lambda_1 M_1 , \frac{\mathbb 1}{d_{I'}}\right) =& \mu \left(\frac{\mathbb 1}{d_I}  + \lambda_1 M_1 \frac{\mathbb 1}{d_{I'}} \right) = \left(\frac{\mathbb 1}{d_I}  + \lambda_1 M_1\right)\otimes \frac{\mathbb 1}{d_{I'}}  = \frac{1}{d_I d_{I'}} \mathbb 1 \otimes \mathbb 1 + \frac{\lambda_1}{d_{I'}} M_1 \otimes \mathbb 1 \nonumber \\ 
=& \mu\left(\frac{\mathbb 1}{d_I}, \frac{\mathbb 1}{d_{I'}}\right) + \frac{\lambda_1}{d_{I'}} M_1 \otimes \mathbb 1 \nonumber =  \mu^L\left(\frac{\mathbb 1}{d_I}, \frac{\mathbb 1}{d_{I'}}\right)  +\frac{\lambda_1}{d_{I'}} M_1 \otimes \mathbb 1.
\end{align}
Therefore by bilinearity we find $\mu^L(M_1,\mathbb 1) = M_1 \otimes \mathbb 1$ and similarly $\mu^L(\mathbb 1, M_2) = \mathbb 1 \otimes M_2$. Similarly, applying R.2 to the process matrices $ \frac{\mathbb 1}{d_I}  + \lambda_1 M_1$ and $\frac{\mathbb 1}{d_{I'}}  + \lambda_2 M_2$, which have the same signaling direction, we obtain
\begin{align}
\mu^L\left(\frac{\mathbb 1}{d_I}  + \lambda_1 M_1,  \frac{\mathbb 1}{d_{I'}}  + \lambda_2 M_2\right) = \mu \left(\frac{\mathbb 1}{d_I}  + \lambda_1 M_1,  \frac{\mathbb 1}{d_{I'}}  + \lambda_2 M_2\right) = \left(\frac{\mathbb 1}{d_I} + \lambda_1 M_1\right)\otimes \left( \frac{\mathbb 1}{d_{I'}} + \lambda_2 M_2\right).
\label{eq:mu_L_MM}
\end{align}

Collecting our results and using bilinearity on the left hand side of Eq.~\eqref{eq:mu_L_MM} above, we finally see that R'.2 is satisfied:
\begin{equation}
\mu^L(M_1,M_2) =M_1 \otimes M_2.
\end{equation}

\section{Proof of Th.\ref{th1} in arbitrary dimension}\label{app:B}
In this appendix, we will extend the proof of Th.~\ref{th1} to the case of arbitrary dimension. We start with the following

\begin{lemma}\label{lemma:eig_PTI}
Let $M\in {\rm PTI}_{\rm AB}$ and $N\in {\rm PTI}_{\rm A'B'}$ be Hermitian operators such that $\ket{k}=\ket{k_1}_\Ain\otimes \ket{k_2}_\Aou \otimes \ket{k_3}_\Bin \otimes\ket{k_4}_\Bou$ is an eigenvector for $M$, with eigenvalues given, according to the above factorization, by the products $\lambda_k=\lambda_k^{(1)}\lambda_k^{(2)}\lambda_k^{(3)}\lambda_k^{(4)}$, with $\lambda_k^{(i)}\in\{-1,0,1\}$ $i=1,2,3,4$, and, similarly,  ${\ket{j}=\ket{j_1}_{\Ain'}\otimes \ket{j_2}_{\Aou'} \otimes \ket{j_3}_{\Bin'} \otimes\ket{j_4}_{\Bou'}}$ is an eigenvector of $N$, with eigenvalue $\eta_j=\eta_j^{(1)}\eta_j^{(2)}\eta_j^{(3)}\eta_j^{(4)}$, with $\eta_j^{(i)}\in\{-1,0,1\}$ $i=1,2,3,4$. We then have
\begin{equation}
\mu(M,N)\ket{k,j}=\lambda_k \eta_j \ket{k,j}.
\end{equation}
\end{lemma} 
{\it Proof.---}  The case $\lambda_k,\eta_j=\pm 1$ are included in Lemma 2. Let us consider the case $M\ket{k}=0$ and $N\ket{j}\neq 0$, the case $M\ket{k}=N\ket{j}= 0$ can be obtained in a similar way, by applying the same argument first to $M$, then to $N$. Since $M$ is in ${\rm PTI}_{\rm AB}$, we can write it as $M=X^1_\Ain\otimes X^2_\Aou \otimes X^3_\Bin \otimes X^4_\Bou$. Let us now further assume $X^1\ket{k_1}_{\Ain}=0$, and $X^i\ket{k_i}_{\rm Y}\neq 0$ for $i=2,3,4$, ${\rm Y}=\Aou,\Bin,\Bou$, in particular, this implies that $\ket{k_i}$ are eigenvectors for eigenvalues $\pm 1$ for $i=2,3,4$. We can then write:
\begin{equation}\label{eq:split_0ev}
X^1 = \Big(X^1 + \ketbra{k_1} - \ketbra{(k+1)_1}\Big) + \Big(\ketbra{(k+1)_1} - \ketbra{k_1}\Big)=: X'^1+X''^1,
\end{equation}
where $\ket{(k+1)_1}$ is a vector orthogonal to $\ket{k_1}$. Then $X'^1,X''^1$  are both traceless and $X'^1\ket{k_1}=\ket{k_1}$, $X''^1\ket{k_1}=-\ket{k_1}$. We then have that $M':= X'^1_\Ain\otimes X^2_\Aou \otimes X^3_\Bin \otimes X^4_\Bou$ and $M'':= X''^1_\Ain\otimes X^2_\Aou \otimes X^3_\Bin \otimes X^4_\Bou$ are again in ${\rm PTI}_{\rm AB}$. Thus, by Lemma 2, 
\begin{equation}
\mu(M,N)\ket{k,j}=\mu(M'+M'',N)\ket{k,j} = \mu(M',N)\ket{k,j} + \mu(M'',N)\ket{k,j} = M'\otimes N\ket{k,j} + M''\otimes N\ket{k,j}=0. 
\end{equation}
If another operator, say $X^2$, is zero on the corresponding eigenvector, say $\ket{k_2}_{\Aou}$, we can again repeat the construction in Eq.~\eqref{eq:split_0ev} to construct  $X'^2,X''^2$ with $+1,-1$ eigenvalues and use again linearity and Lemma 2. 
Similarly, the same argument can be extended to all $X^i$ and to $N$. \qed

To conclude the proof of Th.~\ref{th1}, it is sufficient to construct a basis of operators containing the identity and where each elements, except the identity, is traceless and with eigenvalues in $\{-1,0,1\}$. Let $\mathcal{H}$ be a Hilbert space with dimension $d$, and let $\{ \ket{k} \}_{k = 1}^{d}$ be a basis for $\HH$. The space of Hermitian operators on $\HH$ is a real vector space of dimension $d^2$. We define the following operators
\begin{align}
Z_i &= |i\rangle \langle i | - |i + 1\rangle \langle i+1|, \quad 1 \leq i \leq d -1 \\
X_{jk} &= |j \rangle \langle k| + |k \rangle \langle j|, \quad 1 \leq j < k \leq d \\
Y_{jk} &= i (|j \rangle \langle k| - |k\rangle \langle j|), \quad 1 \leq j < k \leq d,
\end{align} 
which are traceless, hermitian and with eigenvalues in $\{-1,0,1\}$. The $X_{jk}$ and $Y_{jk}$ are also known as part of an operator basis called \textit{Generalized Gell-Mann matrices}~\cite{BlochVectorsQudits}. For completeness we now show that, together with $\openone$, the above set of matrices form a basis for the space of Hermitian operators on $\mathcal{H}$. It is clear that the $\{X_{jk} \}$ and $\{Y_{jk} \}$ span the space of Hermitian operators whose diagonal is zero in the $| k \rangle$ basis. All that remains to be shown is that $\{ \openone, Z_i \}$ forms a basis for the space of diagonal real matrices, which we prove by expressing the basis $\{|k \rangle \langle k|\}$ in terms of the new basis $\{ \openone, Z_i \}$.

Notice that for $1 \leq i \leq d -1$,
\begin{equation}
\sum_{j = i}^{j = d - 1} Z_j = |i \rangle \langle i| - |d \rangle \langle d|,
\end{equation}
and also that
\begin{align}
\sum_{j=1}^{d-1} j Z_j = \sum_{j=1}^{d-1} |j \rangle \langle j| - (d-1) |d \rangle \langle d| = \openone - d |d\rangle \langle d|.
\end{align}
Combining the above two expressions gives
\begin{align}
|d \rangle \langle d| &= \frac{1}{d} \openone - \frac{1}{d} \sum_{j=1}^{d-1} j Z_j, \\
| i \rangle \langle i | &= \frac{1}{d} \openone  + \sum_{j= i}^{d-1} Z_j - \frac{1}{d} \sum_{j=1}^{d-1} j Z_j, \quad 1 \leq i < d,
\end{align}
which concludes that $\{ \openone, Z_i, X_{jk}, Y_{jk} \}$ is a basis for the space of Hermitian operators of $\mathcal{H}$.

We can use the above construction to build a basis for $\mathcal{L}(\mathcal{H}_{A_I}\otimes \mathcal{H}_{A_O} \otimes \mathcal{H}_{B_I} \otimes \mathcal{H}_{B_O})$ consisting of tensor products of local Hermitian operators whose eigenvalues are in $\{-1,0,1\}$. We then remove from this basis all the terms that do not satisfy the linear constraints $L_V$. This gives us a basis for the linear space of valid $W$s, which is contained in ${\rm PTI}_{\rm AB}$. We will call this basis simply $\{ M_i\}_{i\in I}$, and by Lemma~\ref{lemma:eig_PTI}, we have
\begin{equation}
\mu(M_i,M_j)=M_i\otimes M_j.
\end{equation}

We can then decompose any pair $W,W'$ as 
\begin{equation}
W = \openone + \sum_i c_i M_i , \quad W' = \openone + \sum_i d_i M_i ,
\end{equation} 
and apply $\mu$, namely
\begin{equation}
\begin{split}
\mu(W,W')&=\openone + \sum_i c_i M_i\otimes \openone + \openone\otimes \sum_i d_i M_i + \sum_{ij}  c_i d_j  \mu(M_i, M_j)  \\
&=\openone + \sum_i c_i M_i\otimes \openone + \openone\otimes \sum_i d_i M_i + \sum_{ij}  c_i d_j M_i\otimes M_j  
= W\otimes W',
\end{split}
\end{equation}
which concludes the proof of Th.~\ref{th1}.

\providecommand{\href}[2]{#2}\begingroup\raggedright\endgroup

\end{widetext}


\begin{thebibliography}{10}

\bibitem{Bell1964}
J.~S. Bell, ``On the {E}instein--{P}odolsky--{R}osen Paradox,'' {\em Physics}
  {\bfseries 1}, 195 (1964).

\bibitem{FineTuningBellSpekkens}
C.~J. {Wood} and R.~W. {Spekkens}, ``{The lesson of causal discovery algorithms
  for quantum correlations: causal explanations of Bell-inequality violations
  require fine-tuning},''
  \href{http://dx.doi.org/10.1088/1367-2630/17/3/033002}{{\em New J. Phys.}
  {\bfseries 17}, 033002 (2015)},
  \href{http://arxiv.org/abs/1208.4119}{{\ttfamily arXiv:1208.4119
  [quant-ph]}}.

\bibitem{Hardy2005}
L.~Hardy, ``Probability Theories with Dynamic Causal Structure: A New Framework
  for Quantum Gravity,'' \href{http://arxiv.org/abs/gr-qc/0509120}{{\ttfamily
  arXiv:gr-qc/0509120 [gr-qc]}}.

\bibitem{Hardy2007}
L.~Hardy, ``Towards quantum gravity: a framework for probabilistic theories
  with non-fixed causal structure,''
  \href{http://dx.doi.org/10.1088/1751-8113/40/12/S12}{{\em J. Phys. A: Math.
  Theor.} {\bfseries 40}, 3081 (2007)},
  \href{http://arxiv.org/abs/gr-qc/0608043}{{\ttfamily arXiv:gr-qc/0608043
  [gr-qc]}}.

\bibitem{Hardy2009}
L.~Hardy, \href{http://dx.doi.org/10.1007/978-1-4020-9107-0_21}{``Quantum
  Gravity Computers: On the Theory of Computation with Indefinite Causal
  Structure,''} in {\em Quantum Reality, Relativistic Causality, and Closing
  the Epistemic Circle: Essays in Honour of Abner Shimony}, pp.~379--401.
\newblock Springer Netherlands, Dordrecht, 2009.
\newblock \href{http://arxiv.org/abs/quant-ph/0701019}{{\ttfamily
  arXiv:quant-ph/0701019 [quant-ph]}}.

\bibitem{Oreshkov2012}
O.~Oreshkov, F.~Costa, and {\v{C}}.~Brukner, ``Quantum correlations with no
  causal order,'' \href{http://dx.doi.org/10.1038/ncomms2076}{{\em Nat.
  Commun.} {\bfseries 3}, 1092 (2012)},
  \href{http://arxiv.org/abs/1105.4464}{{\ttfamily arXiv:1105.4464
  [quant-ph]}}.

\bibitem{Chiribella2013}
G.~Chiribella, G.~M. D'Ariano, P.~Perinotti, and B.~Valiron, ``Quantum
  computations without definite causal structure,''
  \href{http://dx.doi.org/10.1103/PhysRevA.88.022318}{{\em Phys. Rev. A}
  {\bfseries 88}, 022318 (2013)},
  \href{http://arxiv.org/abs/0912.0195}{{\ttfamily arXiv:0912.0195
  [quant-ph]}}.

\bibitem{Oeckl2016}
R.~Oeckl, ``A local and operational framework for the foundations of physics,''
  \href{http://arxiv.org/abs/1610.09052}{{\ttfamily arXiv:1610.09052
  [quant-ph]}}.

\bibitem{Araujo2015}
M.~Ara{\'u}jo, C.~Branciard, F.~Costa, A.~Feix, C.~Giarmatzi, and
  {\v{C}}.~Brukner, ``Witnessing causal nonseparability,''
  \href{http://dx.doi.org/10.1088/1367-2630/17/10/102001}{{\em New J. Phys.}
  {\bfseries 17}, 102001 (2015)},
  \href{http://arxiv.org/abs/1506.03776}{{\ttfamily arXiv:1506.03776
  [quant-ph]}}.

\bibitem{OpticalQuantumSwitchProcopio}
L.~M. {Procopio}, A.~{Moqanaki}, M.~{Ara{\'u}jo}, F.~{Costa}, I.~{Alonso
  Calafell}, E.~G. {Dowd}, D.~R. {Hamel}, L.~A. {Rozema}, {\v C}.~{Brukner},
  and P.~{Walther}, ``{Experimental superposition of orders of quantum
  gates},'' \href{http://dx.doi.org/10.1038/ncomms8913}{{\em Nat. Commun.}
  {\bfseries 6}, 7913 (2015)}, \href{http://arxiv.org/abs/1412.4006}{{\ttfamily
  arXiv:1412.4006 [quant-ph]}}.

\bibitem{OpticalQuantumSwitchRubino}
G.~Rubino, L.~A. Rozema, A.~Feix, M.~Ara{\'u}jo, J.~M. Zeuner, L.~M. Procopio,
  {\v C}.~Brukner, and P.~Walther, ``Experimental verification of an indefinite
  causal order,'' \href{http://dx.doi.org/10.1126/sciadv.1602589}{{\em Sci.
  Adv.} {\bfseries 3}, (2017)},
  \href{http://arxiv.org/abs/1608.01683}{{\ttfamily arXiv:1608.01683
  [quant-ph]}}.

\bibitem{Goswami2018}
K.~Goswami, C.~Giarmatzi, M.~Kewming, F.~Costa, C.~Branciard, J.~Romero, and
  A.~G. White, ``Indefinite Causal Order in a Quantum Switch,''
  \href{http://arxiv.org/abs/1803.04302}{{\ttfamily arXiv:1803.04302
  [quant-ph]}}.

\bibitem{Oreshkov2018}
O.~Oreshkov, ``{On the whereabouts of the local operations in physical
  realizations of quantum processes with indefinite causal order},''
  \href{http://arxiv.org/abs/1801.07594}{{\ttfamily arXiv:1801.07594}}.

\bibitem{Oreshkov2016}
O.~Oreshkov and C.~Giarmatzi, ``Causal and causally separable processes,''
  \href{http://dx.doi.org/10.1088/1367-2630/18/9/093020}{{\em New Journal of
  Physics} {\bfseries 18}, 093020 (2016)},
  \href{http://arxiv.org/abs/1506.05449}{{\ttfamily arXiv:1506.05449
  [quant-ph]}}.

\bibitem{Branciard:2016aa}
C.~Branciard, M.~Ara\'{u}jo, A.~Feix, F.~Costa, and {\v{C}}.~Brukner, ``The
  simplest causal inequalities and their violation,''
  \href{http://dx.doi.org/10.1088/1367-2630/18/1/013008}{{\em New J. Phys.}
  {\bfseries 18}, 013008 (2016)},
  \href{http://arxiv.org/abs/1508.01704}{{\ttfamily arXiv:1508.01704
  [quant-ph]}}.

\bibitem{Abbott:2016aa}
A.~A. Abbott, C.~Giarmatzi, F.~Costa, and C.~Branciard, ``Multipartite causal
  correlations: Polytopes and inequalities,''
  \href{http://dx.doi.org/10.1103/PhysRevA.94.032131}{{\em Phys. Rev. A}
  {\bfseries 94}, 032131 (2016)},
  \href{http://arxiv.org/abs/1608.01528}{{\ttfamily arXiv:1608.01528
  [quant-ph]}}.

\bibitem{Miklin_ineq:2017}
N.~Miklin, A.~A. Abbott, C.~Branciard, R.~Chaves, and C.~Budroni, ``The
  entropic approach to causal correlations,''
  \href{http://dx.doi.org/10.1088/1367-2630/aa8f9f}{{\em New J. Phys.}
  {\bfseries 19}, 113041 (2017)},
  \href{http://arxiv.org/abs/1608.01528}{{\ttfamily arXiv:1608.01528
  [quant-ph]}}.

\bibitem{AraujoPRL2014}
M.~Ara\'ujo, F.~Costa, and {\v{C}}.~Brukner, ``Computational Advantage from
  Quantum-Controlled Ordering of Gates,''
  \href{http://dx.doi.org/10.1103/PhysRevLett.113.250402}{{\em Phys. Rev.
  Lett.} {\bfseries 113}, 250402 (2014)}.

\bibitem{Feix2015}
A.~Feix, M.~Ara\'{u}jo, and {\v{C}}.~Brukner, ``{Quantum superposition of the
  order of parties as a communication resource},''
  \href{http://dx.doi.org/10.1103/PhysRevA.92.052326}{{\em Phys. Rev. A}
  {\bfseries 92}, 052326 (2015)},
  \href{http://arxiv.org/abs/1508.07840}{{\ttfamily arXiv:1508.07840}}.

\bibitem{AllardPRL2016}
P.~A.~Gu\'erin, A.~Feix, M.~Ara\'ujo, and {\v{C}}.~Brukner, ``Exponential
  Communication Complexity Advantage from Quantum Superposition of the
  Direction of Communication,''
  \href{http://dx.doi.org/10.1103/PhysRevLett.117.100502}{{\em Phys. Rev.
  Lett.} {\bfseries 117}, 100502 (2016)},
  \href{http://arxiv.org/abs/1605.07372}{{\ttfamily arxiv:1605.07372
  [quant-ph]}}.

\bibitem{Baumeler2017}
{\"{A}}.~Baumeler and S.~Wolf, ``Non-Causal Computation,''
  \href{http://dx.doi.org/10.3390/e19070326}{{\em Entropy} {\bfseries 19},
  (2017)}, \href{http://arxiv.org/abs/1601.06522}{{\ttfamily
  arXiv:1601.06522}}.

\bibitem{Araujo2017_CTC}
M.~Ara\'ujo, P.~A. Gu\'erin, and A.~Baumeler, ``Quantum computation with
  indefinite causal structures,''
  \href{http://dx.doi.org/10.1103/PhysRevA.96.052315}{{\em Phys. Rev. A}
  {\bfseries 96}, 052315 (2017)},
  \href{http://arxiv.org/abs/1706.09854}{{\ttfamily arXiv:1706.09854}}.

\bibitem{Baumeler2018}
{\"{A}}.~Baumeler and S.~Wolf, ``{Computational tameness of classical
  non-causal models},'' \href{http://dx.doi.org/10.1098/rspa.2017.0698}{{\em
  Proc. R. Soc. A} {\bfseries 474}, (2018)},
  \href{http://arxiv.org/abs/1611.05641}{{\ttfamily arXiv:1611.05641}}.

\bibitem{Ebler2018}
D.~Ebler, S.~Salek, and G.~Chiribella, ``Enhanced Communication with the
  Assistance of Indefinite Causal Order,''
  \href{http://dx.doi.org/10.1103/PhysRevLett.120.120502}{{\em Phys. Rev.
  Lett.} {\bfseries 120}, 120502 (2018)},
  \href{http://arxiv.org/abs/1711.10165}{{\ttfamily arxiv:1711.10165
  [quant-ph]}}.

\bibitem{Nielsen2002}
M.~A. Nielsen and I.~L. Chuang, {\em Quantum Computation and Quantum
  Information: 10th Anniversary Edition}.
\newblock Cambridge University Press, New York, NY, USA, 10th~ed., 2011.

\bibitem{Wilde2017}
M.~M. Wilde, \href{http://dx.doi.org/10.1017/9781316809976}{{\em Quantum
  Information Theory}}.
\newblock Cambridge University Press, 2~ed., 2017.

\bibitem{SchumacherCompression}
B.~Schumacher, ``Quantum coding,''
  \href{http://dx.doi.org/10.1103/PhysRevA.51.2738}{{\em Phys. Rev. A}
  {\bfseries 51}, 2738--2747 (1995)}.

\bibitem{ZychThesis}
M.~Zych, \href{http://dx.doi.org/10.1007/978-3-319-53192-2}{{\em Quantum
  Systems under Gravitational Time Dilation}}.
\newblock Springer Theses. Springer International Publishing, 2017.

\bibitem{CommonDirectCauseSuperposition}
A.~{Feix} and {\v C}.~{Brukner}, ``{Quantum superpositions of common-cause and
  direct-cause causal structures},''
  \href{http://dx.doi.org/10.1088/1367-2630/aa9b1a}{{\em New J. Phys.}
  {\bfseries 19}, 123028 (2017)},
  \href{http://arxiv.org/abs/1606.09241}{{\ttfamily arXiv:1606.09241
  [quant-ph]}}.

\bibitem{GravityQuantumSwitch}
M.~{Zych}, F.~{Costa}, I.~{Pikovski}, and C.~{Brukner}, ``{Bell's Theorem for
  Temporal Order},'' \href{http://arxiv.org/abs/1708.00248}{{\ttfamily
  arXiv:1708.00248 [quant-ph]}}.

\bibitem{Chiribella2009}
G.~Chiribella, G.~M. D'Ariano, and P.~Perinotti, ``Theoretical framework for
  quantum networks,'' \href{http://dx.doi.org/10.1103/PhysRevA.80.022339}{{\em
  Phys. Rev. A} {\bfseries 80}, 022339 (2009)},
  \href{http://arxiv.org/abs/0904.4483}{{\ttfamily arXiv:0904.4483
  [quant-ph]}}.

\bibitem{QuantumCausalModelsCosta}
F.~{Costa} and S.~{Shrapnel}, ``{Quantum causal modelling},''
  \href{http://dx.doi.org/10.1088/1367-2630/18/6/063032}{{\em New J. Phys.}
  {\bfseries 18}, 063032 (2016)},
  \href{http://arxiv.org/abs/1512.07106}{{\ttfamily arXiv:1512.07106
  [quant-ph]}}.

\bibitem{Jia2018}
D.~Jia and N.~Sakharwade, ``Tensor products of process matrices with indefinite
  causal structure,'' \href{http://dx.doi.org/10.1103/PhysRevA.97.032110}{{\em
  Phys. Rev. A} {\bfseries 97}, 032110 (2018)}.

\bibitem{Jamiolkovski}
A.~Jamiołkowski, ``Linear transformations which preserve trace and positive
  semidefiniteness of operators,''
  \href{http://dx.doi.org/10.1016/0034-4877(72)90011-0}{{\em Rep. Math. Phys.}
  {\bfseries 3}, 275 -- 278 (1972)}.

\bibitem{Choi}
M.-D. Choi, ``Completely positive linear maps on complex matrices,''
  \href{http://dx.doi.org/10.1016/0024-3795(75)90075-0}{{\em Linear Algebra Its
  Appl.} {\bfseries 10}, 285 -- 290 (1975)}.

\bibitem{Note1}
Alternatively, one could define $\protect \mathcal M_{a|x}$ with a global
  transposition $^t$, taken w.r.t. the $\protect \{| ij \delimiter "526930B
  \protect \}_{ij}$ basis, as in Ref.~\cite {Araujo2015}. This allow one to
  write the process matrix associated to a quantum state $\rho $ shared between
  the parties simply as $\rho _I\otimes \protect \openone _O$, instead of $\rho
  ^t_I\otimes \protect \openone _O$. We will not use this convention here.

\bibitem{Perinotti2016}
P.~Perinotti, \href{http://dx.doi.org/10.1007/978-3-319-68655-4}{``Causal
  structures and the classification of higher order quantum computations,''} in
  {\em Time in Physics (Tutorials, Schools, and Workshops in the Mathematical
  Sciences)}, R.~Renner and S.~Stupar, eds.
\newblock Birkh\"auser Basel, 1st~ed., 2017.
\newblock \href{http://arxiv.org/abs/1612.05099}{{\ttfamily arXiv:1612.05099
  [quant-ph]}}.

\bibitem{Bisio2018}
A.~Bisio and P.~Perinotti, ``Axiomatic theory of Higher-Order Quantum
  Computation,'' \href{http://arxiv.org/abs/1806.09554}{{\ttfamily
  arXiv:1806.09554 [quant-ph]}}.

\bibitem{Renner2006}
R.~Renner, S.~Wolf, and J.~Wullschleger,
  \href{http://dx.doi.org/10.1109/ISIT.2006.262081}{``The Single-Serving
  Channel Capacity,''} in {\em 2006 IEEE International Symposium on Information
  Theory}.
\newblock IEEE, 2006.
\newblock \href{http://arxiv.org/abs/cs/0608018}{{\ttfamily arXiv:cs/0608018
  [cs.IT]}}.

\bibitem{Konig2009}
R.~Konig, R.~Renner, and C.~Schaffner, ``The Operational Meaning of Min- and
  Max-Entropy,'' \href{http://dx.doi.org/10.1109/TIT.2009.2025545}{{\em IEEE
  Transactions on Information Theory} {\bfseries 55}, 4337--4347 (2009)},
  \href{http://arxiv.org/abs/0807.1338}{{\ttfamily arXiv:0807.1338
  [quant-ph]}}.

\bibitem{Chiribella2016}
G.~Chiribella and D.~Ebler, ``Optimal quantum networks and one-shot
  entropies,'' \href{http://dx.doi.org/10.1088/1367-2630/18/9/093053}{{\em New
  Journal of Physics} {\bfseries 18}, 093053 (2016)},
  \href{http://arxiv.org/abs/1606.02394}{{\ttfamily arXiv:1606.02394
  [quant-ph]}}.

\bibitem{Hardy2001}
L.~Hardy, ``Quantum theory from five reasonable axioms,''
  \href{http://arxiv.org/abs/quant-ph/0101012}{{\ttfamily
  arxiv:quant-ph/0101012}}.

\bibitem{Barrett2007}
J.~Barrett, ``Information processing in generalized probabilistic theories,''
  \href{http://dx.doi.org/10.1103/PhysRevA.75.032304}{{\em Phys. Rev. A}
  {\bfseries 75}, 032304 (2007)},
  \href{http://arxiv.org/abs/quant-ph/0508211}{{\ttfamily
  arxiv:quant-ph/0508211}}.

\bibitem{Hardy2009_foliable}
L.~Hardy, ``Foliable Operational Structures for General Probabilistic
  Theories,'' \href{http://arxiv.org/abs/0912.4740}{{\ttfamily arXiv:0912.4740
  [quant-ph]}}.

\bibitem{Hardy2011}
L.~Hardy, ``Reformulating and Reconstructing Quantum Theory,''
  \href{http://arxiv.org/abs/1104.2066}{{\ttfamily arXiv:1104.2066
  [quant-ph]}}.

\bibitem{BlochVectorsQudits}
R.~A. {Bertlmann} and P.~{Krammer}, ``{Bloch vectors for qudits},''
  \href{http://dx.doi.org/10.1088/1751-8113/41/23/235303}{{\em J. Phys. A:
  Math. Theor.} {\bfseries 41}, 235303 (2008)},
  \href{http://arxiv.org/abs/0806.1174}{{\ttfamily arXiv:0806.1174
  [quant-ph]}}.

\end{thebibliography}
\end{document}